# Oxygen Diffusion Pathways in Brownmillerite SrCoO$_{2.5}$: Influence of Structure and Chemical Potential

Chandrima Mitra, Tricia Meyer, Ho Nyung Lee, and Fernando A. Reboredo

Materials Science and Technology Division, Oak Ridge National Laboratory, Oak Ridge, Tennessee, 37831, USA

**Abstract**

To design and discover new materials for next-generation energy materials such as solid-oxide fuel cells (SOFCs), a fundamental understanding on their ionic properties and behaviors is essential. The potential applicability of a material for SOFCs is critically determined by the activation energy barrier of oxygen along various diffusion pathways. In this work, we investigate interstitial-oxygen (O$_i$) diffusion in brownmillerite oxide SrCoO$_{2.5}$, employing a first-principles approach. Our calculations indicate highly anisotropic ionic diffusion pathways, which result from its anisotropic crystal structure. The one-dimensional-ordered oxygen vacancy channels are found to provide the easiest diffusion pathway with an activation energy barrier height of 0.62 eV. The directions perpendicular to the vacancy channels have higher energy barriers for O$_{int}$ diffusion. In addition, we have studied migration barriers for oxygen vacancies that could be present as point defects within the material. This in turn could also facilitate the transport of oxygen. Interestingly, for oxygen vacancies, the lowest barrier height was found to occur within the octahedral layer with an energy of 0.82 eV. Our results imply that interstitial migration would be highly one-dimensional in nature. Oxygen vacancy transport, on the other hand, could preferentially occur in the two-dimensional octahedral plane.

## I. Introduction

Solid-oxide fuel cells (SOFCs) are electrochemical devices that convert chemical energy to electricity and vice-versa [1-3]. However, their high operating temperature (as high as 1,000 °C) results in various unwanted reactions and degradation of the cells. Hence a major challenge lies in discovery of new materials and in implementing design principles for enhanced fuel cell performance at reduced temperatures. Specific components of the fuel cell most affected by temperature are the cathode and the electrolyte. This in turn is determined by the reduction of oxygen on the cathode surface, oxygen diffusion into the cathode, and its subsequent transfer into the electrolyte. The overall reactions can be summarized as a set of three equations as can be seen below [4].

$$\frac{1}{2}O_2 + 2e^-_{(cathode)} \leftrightarrow O^{2-}_{(cathode\ surface)}$$

$$O^{2-}_{(cathode\ surface)} \leftrightarrow O^{2-}_{(cathode\ bulk)} \quad\quad (1)$$

$$O_{(cathode\ bulk)} \leftrightarrow O^{2-}_{(electrolyte)}$$

The first reaction refers to the reduction of oxygen on the surface of the cathode, the second reaction describes transport of oxygen into the cathode and the third describes the transfer of oxygen from the cathode to the electrolyte.

As can be seen from the equation above, the reduction rate of oxygen and its diffusion properties in the cathode play a crucial role in determining the cathode performance of SOFCs. Hence efforts must be driven towards achieving faster oxygen ion diffusion in the search for new materials for efficient SOFCs.

In recent years, complex oxides have emerged as promising candidates for cathode materials in energy storage and generation devices. These materials include oxygen-vacancy-ordered brownmillerites (BM, $ABO_{2.5}$), Ruddlesden-Popper phases ($A_{n+1}B_nO_{3n+1}$), and perovskites ($ABO_3$). Compared to conventional perovskites, BMs have received particular attention due to their unique crystal structure. These materials are derived from the perovskite structure via the introduction of an ordered oxygen vacancy framework, which drastically changes the structural and physical properties (see Figure 1). This structural transition may result in a modification of their ionic and catalytic properties, which serve as basic functionalities in energy materials and devices.

Among the BM compounds, $SrCoO_{2.5}$ (SCO) has recently been the subject of intense experimental investigations as a potential cathode material [5-7]. The valence and spin state of cobalt-based materials are particularly sensitive to the oxygen stoichiometry [8]. This is evident from the fact that the BM-SCO is an antiferromagnetic (AFM) insulator, in sharp contrast to its perovskite counterpart $SrCoO_3$, which is a ferromagnetic (FM) metal [9]. This change originates from the removal of 1/6 oxygen atoms, greatly modifying the crystal structure, and valence and spin states of the cobalt ions [8,10]. As shown in Figure 1, the crystal structure of BM-SCO is orthorhombic ($a_o$=5.5739 Å , $b_o$=5.4697 Å , $c_o$=15.7450 Å) [11,7] and is composed of alternating octahedral ($CO_6$, green polyhedra) and tetrahedral layers ($CO_4$, pink polyhedra) along the c axis. This combination results in one-dimensional (1D) oxygen vacancy channels along the [010] direction that could provide open pathways for ionic diffusion. The electronic and magnetic structure of bulk BM-SCO has been investigated earlier in detail [12]. It is a G-type AFM insulator where the bottom of the conduction bands consists of Co 3d states, while the top of the valence bands consist of hybridized O-2p and Co-3d states.

In a recent experimental study, epitaxial growth of high-quality BM-SCO films on $SrTiO_3$ was achieved [5]. What makes BM-SCO particularly interesting is that under normal growth conditions it is thermodynamically much more stable than its perovskite counterpart, $SrCoO_3$. In fact it has been recently shown that the two phases, $SrCoO_{2.5}$ and $SrCoO_3$, could be reversibly switched under oxidizing/reducing conditions, and more importantly, at a temperature as low as 200 °C [7,8]. This low temperature redox capability makes them ideal candidates for possible applications as cathode materials in SOFCs.

A fast catalytic reaction has recently been demonstrated experimentally within BM-SCO [6]. A 100-fold improvement in the oxygen reduction kinetics was achieved when BM-SCO films were grown in a manner such that the ordered vacancy channels were tilted towards the surface and were therefore open to the surface. This underlines the importance of these ordered vacancy structures for oxygen ion transport and calls for a detailed understanding of their specific roles.

In this work we employ quantum mechanical simulations to shed light on this aspect and to elucidate atomic-scale features, which are often difficult to measure in experimental studies. Our study provides a quantitative picture of the exact mechanisms of interstitial oxygen atoms, $O_i$, and oxygen vacancy diffusion pathways in BM-SCO. The activation energy barriers for different mechanisms are computed from a first-principles approach. The anisotropic crystal structure results in anisotropic diffusion properties for both the $O_i$ and oxygen vacancies.

## II. Computational method

All calculations have been performed within density functional theory employing the Vienna Ab-initio Simulations Package (VASP) code [13]. We have used 2 x 2 x 1 supercells for all calculations containing 144 atoms. Projector-augmented wave pseudopotentials [14] have been used with an energy cut of 600 eV. The valence electrons included for Sr, Co, and O are $4s^2\ 4p^6\ 5s^2$, $4s^2\ 3d^7$, and $2s^2\ 2p^4$, respectively. The activation energy barriers for oxygen ion migration and the intermediate transition states have been computed using the nudged elastic band (NEB) method, implemented in the VASP code [15,16]. Within this method a number of intermediate images are optimized along the reaction path. We use 13 intermediate images for our calculations. The energy barriers have been optimized until the forces on each image were converged to 0.004 eV/Å. To take strong correlations into account, the cobalt $d$ orbitals are treated within the local spin density approximation with Hubbard $U$ corrections [17]. A $U$ value of 7.5 eV was chosen and the electronic structure matched closely to that computed with the Hybrid Scuzeria Ernzerhof (HSE) functional [18].

## III. Results and discussions

The activation energy, $E_a$, is a key quantity in all thermally activated processes. It refers to the energy barrier that a system needs to overcome for the process to take place. The overall process is governed by the Arrhenius equation, formulated as follows:

$$k = C \exp(-E_a/k_B T) \qquad (2)$$

in which the rate constant, $k$, of the chemical reaction at any given temperature, T, is determined by $E_a$. To determine the potential applicability of BM-SCO as a cathode material we primarily focus on obtaining $E_a$ for $O_i$ and oxygen vacancies within BM-SCO. Their activation energies would affect their respective rate constants (in Table II, for instance, we present the ratio of rate constants for oxygen vacancy migration along different crystallographic directions). This outcome would determine the preferential route oxygen would take during its transport from the cathode to the electrolyte.

### A. Oxygen interstitial mobility along the 1D vacancy channel

We first study $O_i$ mobility within one-dimensional oxygen vacancy channels in BM-SCO. We simulate this by allowing the interstitial atom to move along the direction of the channels, indicated by green arrows in Fig. 1. The NEB method computes the intermediate configurations as the oxygen interstitial "hops" from one vacant site to the other. The trajectory of such a hopping mechanism from a site *A* to site *B* is represented in Fig. 2. The interstitial atom is found to follow a curved trajectory around cobalt, which is in agreement with the accepted model in perovskites for oxygen ion migration [19-23]. The NEB method performs a constrained optimization by adding spring forces along the "band" between images and projecting out the component of the force due to the potential perpendicular to the force. We find that the most dominant atomic relaxation is that of the nearby cobalt atoms, which relax towards this interstitial oxygen atom. By computing 13 intermediate configurations, a migration barrier energy of 0.62 eV is predicted (Fig. 3).

Here we would like to briefly comment on the magnetic order of the system while the process of oxygen diffusion takes place. In a previous work [12] we reported the magnetic structure of bulk brownmillerite $SrCoO_{2.5}$ (BM-SCO) in detail and found the magnetic exchange parameters of the order of 20 meV. Therefore, we expect only small variations in the migration barriers of oxygen diffusion even if the magnetic structure does change as the $O_i$ atom diffuses through it. In order to accurately determine the magnetic structure of BM-SCO as an $O_i$ atom diffuses through it one would need to simulate a very large supercell to arrive at the dilute limit. However, within the limitation of the supercell size (144 atoms) in our calculations we have compared total energies (with $O_i$) for three different magnetic structures, FM, G-type AFM and an inter-planar AFM structure (where the Co atoms within the octahedral and tetrahedral layers have AFM interactions while those within the same layers have FM orientation). We find the G-type AFM structure still has the lowest energy. However, since perovskite SrCoO$_3$ is known to be a FM metal there must be certain critical concentration of oxygen where a magnetic transition from the G-type AFM structure to the FM structure takes place. A detailed examination of this is beyond the scope of this work.

### B. Oxygen interstitial mobility perpendicular to the 1D vacancy channel

We next consider the migration path of O$_i$ perpendicular to the direction of the channel (along the [100] direction) within the same plane in the tetrahedral layer as shown inFig. 4. We examine three different transport mechanisms I, II and III. The first one (mechanism I) occurs via a simultaneous "hopping" mechanism. In this process the O$_i$ at site X moves to site Y while another oxygen atom at Y "hops" to site Z, and the phenomenon occurs simultaneously as indicated in Fig. 4. We compare this process to two other mechanisms (II and III) along this direction. In mechanism II, this hopping process is not simultaneous but occurs as two separate processes, where first an O atom moves from site Y to Z and then the O$_i$ hops from site X to the unoccupied site Y. The final process, mechanism III, constitutes a direct transport of O$_i$ from initial position X to the final position Z. The results are presented in Table 1. As can be seen from the table, mechanism I has the lowest energy among the three mechanisms considered. Hence this would be the most favorable form of oxygen ion transport along this direction. Mechanism III requires overcoming the highest barrier and is not likely to occur except at sufficiently high temperatures. Mechanism II, on the other hand, which comprises two separate "hopping" processes, has an activation energy closer to that of Mechanism I. This therefore indicates that oxygen ion transport is likely to take place via breaking and making of Co-O bonds along directions perpendicular to the vacancy channels. However, the barrier values of all these mechanisms are significantly higher than what is observed for the oxygen ion transport along the ordered vacancy channel in BM-SCO. The crystallographic structure of BM-SCO, therefore, makes the oxygen ion transport highly anisotropic in nature, with the transport along the vacancy channel to be the most favorable one.

### C. Oxygen Vacancy Mobility

At higher temperatures, the concentration of intrinsic point defects is expected to increase, thereby providing additional routes for oxygen ion migration. Under reducing conditions and at sufficiently low partial pressure of oxygen, oxygen vacancy defects are likely to be present. Furthermore, the presence of Co enhances the formation of oxygen vacancies, which in turn could assist in O$^{2-}$ migration as had been found in other perovskites [24,25]. Based on this consideration, we also investigate oxygen vacancy induced oxygen ion migration in BM-SCO.

To achieve this, we assume oxygen vacancies to be already present while an oxygen atom hops from one vacancy site to the other. This results in the movement of the vacancy opposite to the direction of the oxygen atom. We consider various migration pathways, like before, along different crystallographic directions, as summarize in **Table 2**. We find that the lowest activation energy barrier of 0.82 eV results for vacancy migration along the [010] direction (the *b*-axis) within the octahedral layer. The energy barrier for transport along the *a*-axis ([100]) within the octahedral plane is slightly higher (0.9 eV) because of the presence of octahedral tilts. This results in a greater travel distance for $O_i$ along [100] direction compared to the [010] and hence it requires a higher energy to diffuse along the [100] direction. A similar trend is found to occur within the tetrahedral layer, although the respective energy barriers are found to be higher than what is observed within the octahedral layer.

We understand the difference in oxygen-vacancy migration barriers within the octahedral and tetrahedral layers by correlating them to the formation energies of oxygen vacancies within these two coordination planes. We find that the total energy is about 0.3 eV higher when the vacancy occupies the tetrahedral layer compared to that when it sits on the octahedral one. This suggests that at a given reference chemical potential oxygen vacancies are more likely to form within the octahedral plane. This in turn results in the lower migration energy barrier for oxygen vacancies. It is interesting to note that a similar trend of faster oxygen vacancy diffusion (within octahedral layers) has been reported in other BM oxides [26]. Our calculations therefore confirm that the migration of both $O_i$ and oxygen vacancies within BM-SCO are anisotropic due to its characteristic structure.

### D. Effect of chemical potential

Chemical potentials influence the formation energy of defects in materials. As mentioned above, in first-principles calculations, it explicitly enters into the expression for computing the formation energies of defects, $E^f[X^q]$, as formulated within the Zhang-Northrup formalism [27]:

$$E^f[X^q] = E_{tot}[X^q] - E_{tot}[bulk] - \sum_i n_i \mu_i \qquad (3)$$

The first and second term on the right hand side of the equation refer to the total energies of the supercell with and without the defect respectively, $n_i$ is the number of atoms of type $i$ that have been added to ($n_i > 0$) or removed ($n_i < 0$) from the supercell in order to form the defect and $\mu_i$ are the corresponding chemical potentials. Ultimately, chemical potentials are determined by the experimental growth conditions. In other words one can vary them in equation (3) to explore different experimental scenarios. In our study, therefore, the formation energies of oxygen $O_i$ and oxygen vacancies ($O_{vac}$) defects would be affected by the choice of chemical potential of oxygen in equation (1). Moreover, the chemical potential is a function of temperature *T* and partial pressure $P_i$ of oxygen as follows :

$$\mu_i(P_i, T) = \mu_i^0(P_0, T_0) - k_B T ln[\left(\frac{T}{T_0}\right)^{5/2} \left(\frac{P_0}{P_i}\right)] \qquad (4)$$

where $\mu_i^0(P_0, T_0)$, is the chemical potential of the i-th constituent at pressure $P_0$ (1 atm) and temperature $T_0$ (298 K). When $\mu_i$ is taken as half of the total energy of an oxygen molecule, the condition is frequently referred to as "oxygen-rich". We find that at this chosen reference potential [28] the computed formation energies of $O_{vac}$ at the octahedral and tetrahedral layers are 4.2 and 4.5 eV respectively. On the other hand the formation energy of an $O_i$, at one of the sites on the vacancy channel, attains a much lower value of 0.4 eV at oxygen-rich conditions. At a particular temperature, the chemical potential of oxygen can be further increased by increasing the partial pressure, $P_i$, as indicated in equation (3). Since the formation energy of $O_i$ along the empty one-dimensional channels is low, at a high enough values of chemical potential of oxygen, $O_i$ could form spontaneously and induce a phase change in the brownmillerite structure. In contrast, the formation energy of oxygen vacancies is rather high, and one would need a very low partial pressure or very high temperature in order for them to occur spontaneously. It should be mentioned that the boundaries of chemical potentials could also be set by the formation of various secondary phases. In essence, therefore, the formation energy of these defects (which are influenced by their chemical potentials) would determine their concentration, which in turn would eventually affect their diffusivity.

### E. Effect of *U*

Finally we would like to make a brief comment on the effect of the Hubbard *U* parameter chosen for our calculations. To see its effect we have repeated our calculations for different values of *U*. We find the qualitative nature of our results to be unchanged. However, the absolute numbers vary upon changing the value of U. In Table 3 we present the activation energy barriers for three different values of *U*. The barrier heights are found to increase with increasing values of *U*. As can be seen from the table, the values change by about 0.1 eV (for both oxygen interstitial and vacancy migration), when *U* is varied from 4 to 7.5 eV.

### IV. Conclusions

In conclusion, we have considered BM-SCO as a potential cathode material for SOFCs. Oxygen transport both via interstitial and vacancy mechanisms are demonstrated along different crystallographic pathways. $O_i$ diffusion through the one-dimensional vacancy channel is found to be the dominant mode of transport with the lowest migration energy, highlighting the importance of controlling the crystallographic direction when designing cathode materials for better ionic conduction. As far as oxygen vacancies are concerned, our formation energy calculations indicate that they are most likely to occur in the plane where Co shares octahedral coordination to oxygen. Migration barriers for oxygen vacancies are also found to be dominant within the octahedral plane in the brownmillerite structure, although the octahedral tilts are found to have some effects on the energetics. Since geometric constraints and electronic structure play an important role in determining the activation energy barriers, future studies should be directed towards investigating the effect of chemical substitution on these barrier heights.

**Acknowledgments**

This work was supported by the US Department of Energy, Basic Energy Sciences, Materials Sciences and Engineering Division.

**Figure Captions**

Fig. 1: (Color online) Crystal structure of the bulk perovskite (left) and BM-SCO (right) where the green arrows indicate the direction of $O_i$ diffusion along one-dimensional-ordered vacancy channels (*b*-axis of the crystal)

Fig. 2: (Color online) Trajectory of $O_i$ migration through the vacancy channel within the tetrahedral layer from site *A* to site *B*. The $O_i$ atom moves towards the cobalt atom during its transport to site *B*.

Fig. 3: (Color online) Computed activation barrier height of $O_i$ as it travels from site *A* to *B* within the vacancy channel.

Fig. 4: (Color online) $O_i$ migration perpendicular to the vacancy channel along the *a* axis.

**Tables**

**Table I**: Computed activation energy for an oxygen interstitial atom along the a-axis where the three mechanisms are indicated as I, II, and III as described in the text.

|  | I | II | III |
|---|---|---|---|
| **Energy (eV)** | 1.8 | 2.1 | 2.9 |

**Table II**: Computed activation barrier for oxygen vacancy migration in BM-SCO along two crystallographic directions on both the cobalt containing octahedral and tetrahedral layers. $K_{oct}/K_{tet}$ refer to the ratio of their rate constants (at 600°C) that can be derived from the Arrhenius Equation [eq. (2) in the text].

|  | [010] (eV) | [100] (eV) |
|---|---|---|
| **Octahedral layer** | 0.82 | 0.9 |
| **Tetrahedral layer** | 0.96 | 2.1 |
| $K_{oct}/K_{tet}$ | 6.469 | $8 \times 10^6$ |

Table III: The variation of the energy barriers for different values of the Hubbard $U$ parameter

|  | $U=4$ | $U=5$ | $U=7.5$ |
|---|---|---|---|
| **$O_i$ diffusion ([010])** | 0.54 | 0.58 | 0.62 |
| **O vacancy diffusion ([100])** | 0.7 | 0.75 | 0.82 |

Figures

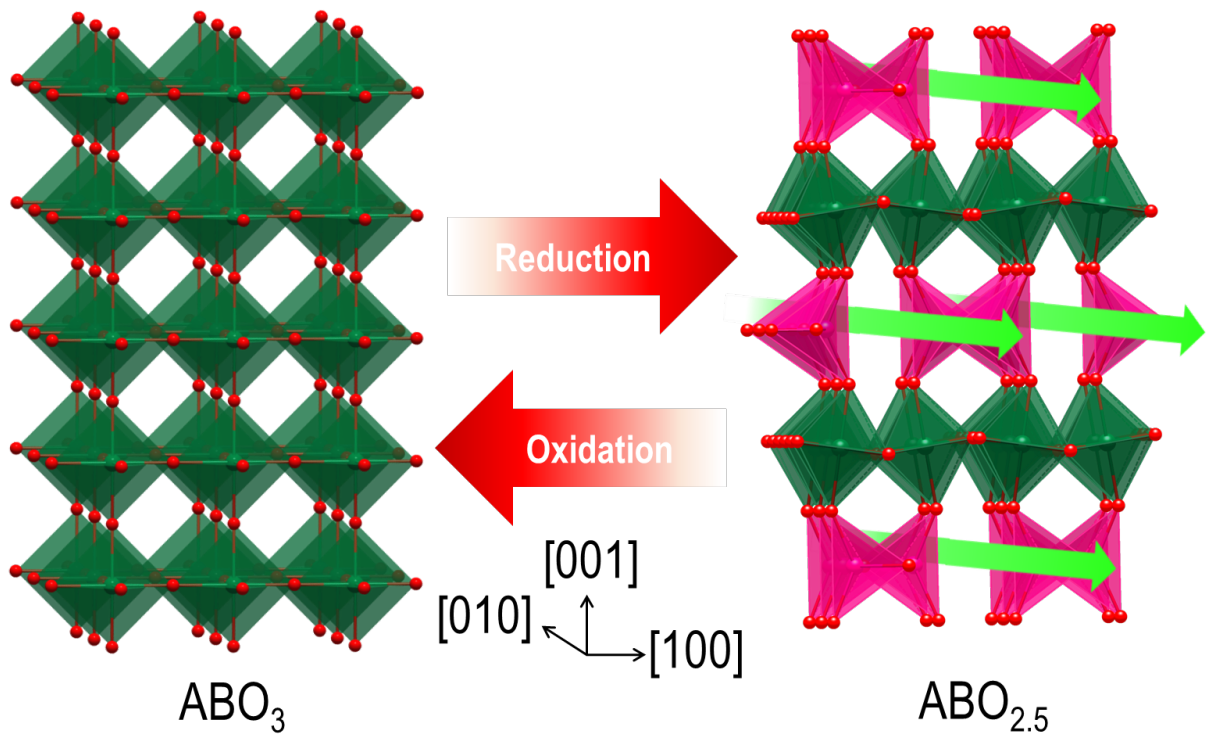

Figure 1

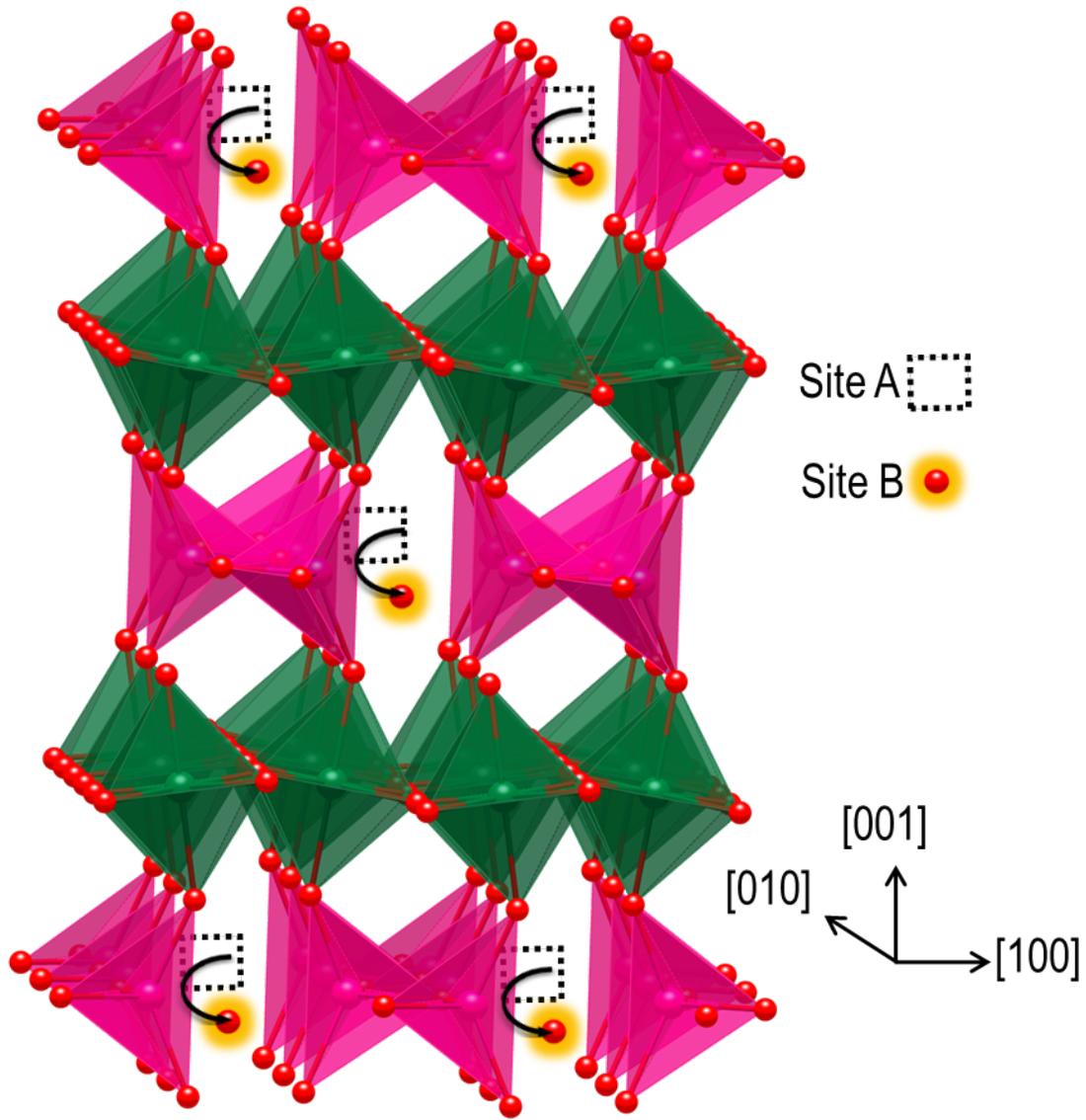

Figure 2

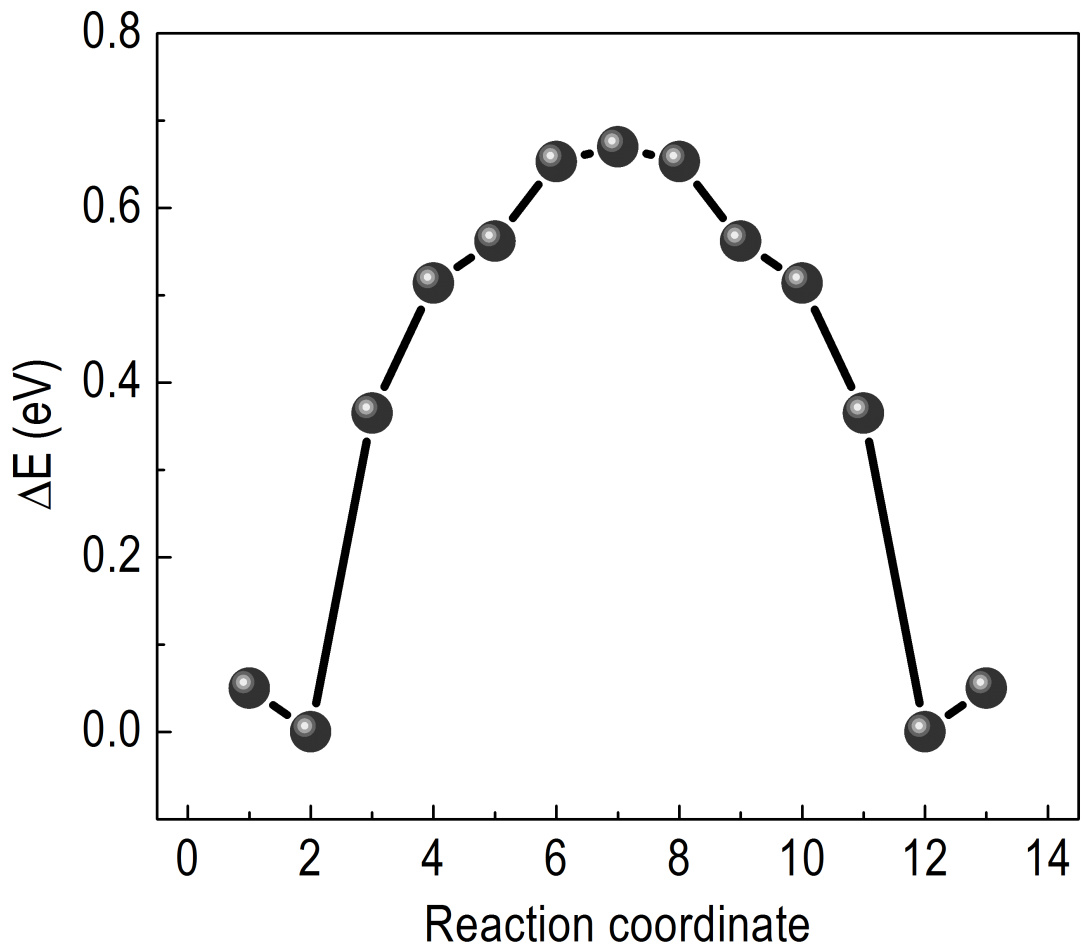

Figure 3

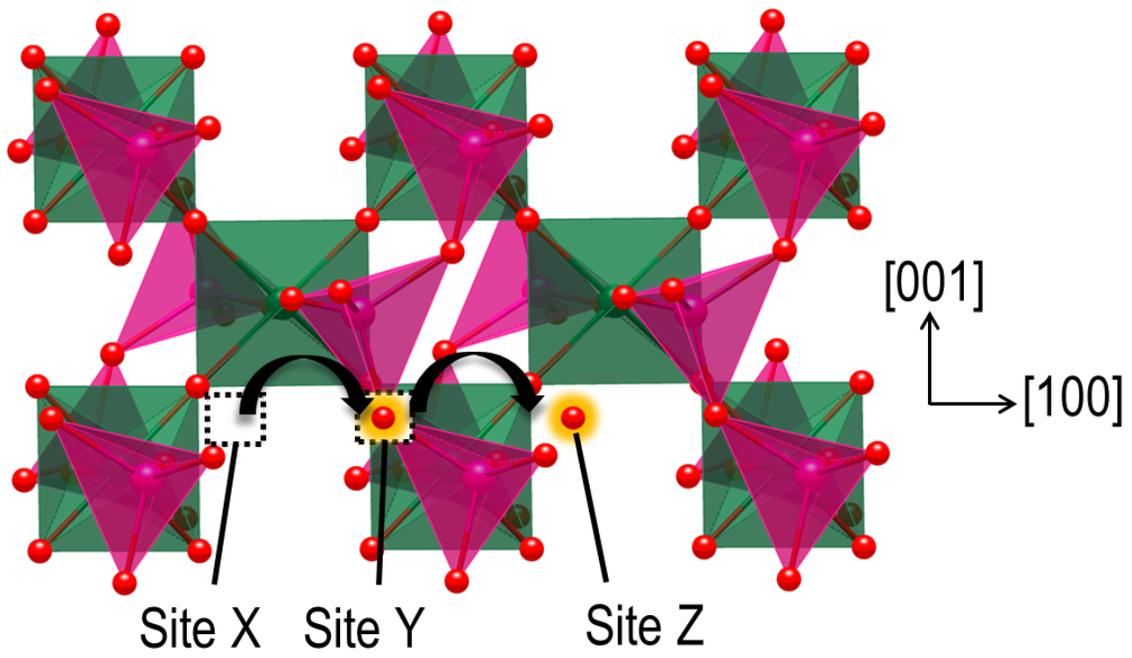

Figure 4